\documentclass[twoside,a4paper,11pt]{sca}
\usepackage{graphicx}
\usepackage{hyperref}
\usepackage{movie15}
\usepackage{natbib}
\newcommand{\msun}{${\rm M}_{\odot}$}

\begin{document}
\pagenumbering{arabic}
\pagestyle{myheadings}
\thispagestyle{empty}
{\flushright\includegraphics[width=\textwidth,bb=90 650 520 700]{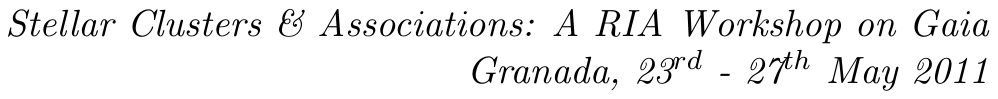}}
\vspace*{0.2cm}
\begin{flushleft}
{\bf {\LARGE
%
The dynamics of star-forming regions -- which mechanisms set the cluster formation efficiency?
%
}\\
\vspace*{1cm}
%
J.~M.~Diederik~Kruijssen$^{1,2,3}$
%
}\\
\vspace*{0.5cm}
%
$^{1}$Astronomical Institute, Utrecht University, PO Box 80000, 3508 TA Utrecht, The Netherlands, {\tt kruijssen@astro.uu.nl};\\
$^{2}$Leiden Observatory, Leiden University, PO Box 9513, 2300 RA Leiden, The Netherlands;\\
$^{3}$Current address: Max-Planck Institut f\"{u}r Astrophysik, Karl-Schwarzschild Stra\ss e 1, 85748 Garching, Germany\\
%
\end{flushleft}
%
\markboth{
Dynamics of star-forming regions
}{ 
%
Diederik~Kruijssen
%
}
\thispagestyle{empty}
\vspace*{0.4cm}
\begin{minipage}[l]{0.09\textwidth}
\ 
\end{minipage}
\begin{minipage}[r]{0.9\textwidth}
\vspace{1cm}
\section*{Abstract}{\small
%
The fraction of star formation that results in bound stellar clusters (cluster formation efficiency or CFE) is a central quantity in many studies of star formation, star clusters and galaxies. Recent results suggest that contrary to popular assumption, the CFE is not (solely) set by gas expulsion, but is also influenced by the primordial environment, although its precise behaviour remains unknown. Here it is discussed which mechanisms set the CFE, which recent advancements have been made to disentangle their contributions, and which studies are needed in the near future to achieve a quantitative understanding of the CFE.
%
\normalsize}
\end{minipage}
%
%
%
\section{Introduction \label{sec:intro}}
Star clusters are popular tracers of the star formation process, both on stellar and galactic scales \citep[e.g.][]{smith07,pflamm07,bastian10}. This approach assumes a certain fraction of star formation that produces bound stellar clusters, also dubbed the {\it cluster formation efficiency} (CFE). It has recently become clear that the CFE is not the product of two different `modes' of dispersed and clustered star formation. Instead, star formation proceeds according to a continuous density spectrum, of which the high-density end yields bound star clusters and the low-density end consists of unbound associations \citep{bressert10,gieles11}. This suggests the existence of a (possibly varying) critical density that allows for the formation of bound structure. If the density spectrum of star formation would vary with environment, this would also imply that the CFE is environmentally dependent -- dense star-forming regions should then yield a high CFE (see e.g. \citealt{adamo11} for a discussion).

It was originally thought that {\it infant mortality} (disruption by gas expulsion) would be the key mechanism behind a CFE lower than 100\% \citep{lada03}. The question thus arises how this new picture, in which stars do not form in clusters but in a hierarchical setting, connects to the concept of infant mortality, and also which other (environmental) processes may determine whether stellar structure ends up being bound. The following sections address the different mechanisms at play, and how Gaia can be used to estimate their relative contributions to the CFE.

\section{Infant mortality: internal disruption by gas expulsion}
In the classical picture of cluster formation, the removal of gas from the star-forming environment by stellar winds and supernovae leads to a perturbation of the gravitational potential. If the gas fraction is large enough, this unbinds the cluster -- the aforementioned `infant mortality' \citep[e.g.][]{bastian06b,goodwin06}. The indications for infant mortality are largely empirical. Surveys of young stellar clusters show that some 90\% of all gas-embedded clusters does not survive the transition to the exposed phase \citep{lada03}. However, this need not imply that the decrease of the number of clusters is causally related to gas expulsion.

As is shown by \citet{bressert10} and \citet{gieles11}, not all stars form in clusters. At least a certain fraction of gas-embedded stellar structure is gravitationally unbound from the onset. While these unbound associations disperse on a crossing time, the bound part of the embedded structure could still be subject to infant mortality. The disruption of a cluster by gas expulsion requires a sufficiently massive central concentration of gas. The theoretical evidence for infant mortality is largely based on analytical estimates \citep[e.g.][]{tutukov78,hills80} or numerical experiments that obey such conditions -- either by assuming a static gas potential\footnote{Except for a normalisation of the gas potential that decreases with time when the gas is expelled.} or dynamical equilibrium between the gas and stars \citep[e.g.][]{boily03b,baumgardt07}.

\begin{figure}
\center
\includegraphics[scale=0.65]{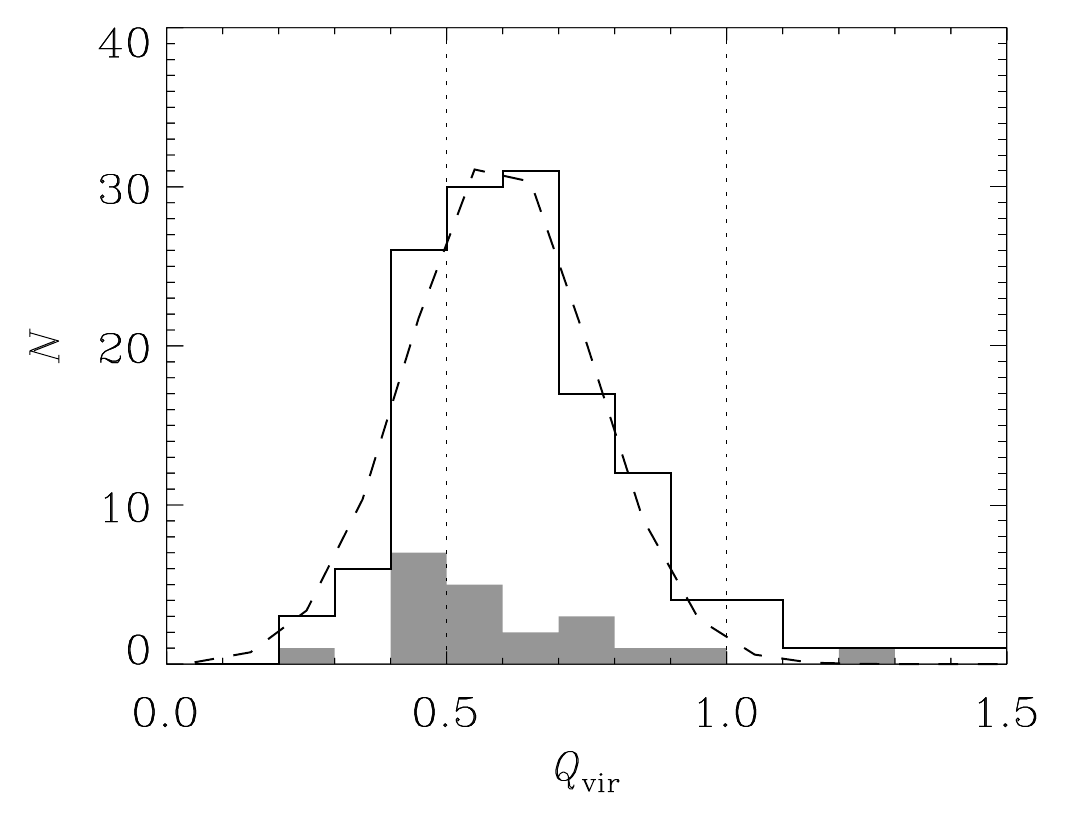}
\caption{\label{fig:virial}
Histogram of the virial ratios $Q_{\rm vir}$ of the subclusters from all snapshots of the simulation (solid line). The shaded histogram represents the set of subclusters from the last snapshot at $t=0.641$~Myr (after approximately one free-fall time). The dashed line is a Gaussian fit to the data for all snapshots, with mean value $Q_{\rm vir}=0.59$ and standard deviation $\sigma_Q=0.16$. The vertical dotted lines again indicate the marginally gravitationally bound case ($Q_{\rm vir}<1$) and the virialised case ($Q_{\rm vir}=0.5$).
}
\end{figure}

It was recently found by \citet{offner09} that the velocity dispersion of stars in simulations of star formation are about a factor of five smaller than that of the gas. This result has recently been investigated further by \citet{kruijssen11c}, where we analysed the dynamics of the stellar structure in the simulation by \citet{bonnell08}. They identify subclusters using a minimum spanning tree \citep[see][]{maschberger10} and compute the virial ratios of the stellar component {\it only}, by ignoring the gravitational potential of the gas. This is equivalent to observing the system at the moment of instantaneous gas expulsion. The virial ratios are also corrected for binaries and higher-order multiple systems. The resulting distribution of virial ratios is shown in Fig.~\ref{fig:virial} for the ensemble population of subclusters from all snapshots. On average, the subclusters are very close to virial equilibrium when neglecting the gas, particularly after one free-fall time has passed.

The virialised state of the subclusters is traced to the fact that they are generally gas-poor, with typical gas fractions below 10\% within their half-mass radii. Since the simulation by \citet{bonnell08} does not include feedback by radiation or stellar winds, the low gas fractions must have a dynamical origin. About half of the gas depletion can be attributed to the accretion of gas onto the sink particles, while the other half is due to the accretion-induced shrinkage of the subclusters, which is the dynamical response of the subcluster to the mass increase of the sink particles \citep[see][]{moeckel11}. This indicates that the accretion of gas onto the sinks is sufficient to balance the overall gas inflow. The subclusters only become gas-rich (with gas fractions above 20\%) beyond about three half-mass radii, and are thus embedded in an evacuated cocoon of gas. The lack of a central concentration of gas implies that the subclusters are relatively unperturbed by gas expulsion -- a simple analytical estimate yields an $\sim 8$\% expansion after the gas has been removed.

The degree to which the subclusters are able to evacuate the surrounding gas dynamically depends on the free-fall time. For a certain density spectrum of star formation, a gas-poor state is more easily achieved in the high-density regions, which complete the largest number of free-fall times before the onset of gas expulsion. In low-density regions, the free-fall time is longer and the evacuation would be less efficient, implying that the disruptive effect of infant mortality could be more important. It is worth noting that the densities at which the effect of infant mortality is strongest thus coincide with the range where stars are predominantly formed in unbound associations.

\section{The cruel cradle effect: external disruption by the primordial environment}
The transition of young star clusters from the gas-embedded to the exposed phase is also accompanied by an environmental effect that decreases the number of clusters more strongly before and during gas expulsion than after it. Star clusters are tidally disrupted by passing giant molecular clouds (GMCs), particularly in dense environments, where the frequency and strength of the tidal shocks is high \citep{gieles06}. In recent work, it has been indicated that star-forming regions affect their offspring in this way: the GMCs in these regions are capable of efficiently disrupting the new-born stellar clusters \citep{elmegreen10b,kruijssen11}. In those cases where the ambient density is high enough, the tidal shocks would be capable of disrupting young clusters in a single encounter, independently of their mass \citep[also see][]{gieles06}. This form of enhanced disruption is most prevalent in star-forming regions because the mean disruption rate decreases as a cluster population ages. We identified the two responsible mechanisms for a decreasing disruption rate in \citet[see their Fig.~10]{kruijssen11}: {\it cluster migration}, i.e. the motion of clusters away from dense star-forming regions into the field, and {\it natural selection}, i.e. the preferential survival of those clusters residing in less disruptive environments. Because the enhanced disruption terminates when the gas has been expelled, it influences the cluster population in a way that is very similar to infant mortality. However, rather than being an internal effect like infant mortality is, the primordial disruption by tidal shocks is external. In \citet{kruijssen11c}, we therefore named it the {\it cruel cradle effect}.

\begin{figure}
\center
\includegraphics[scale=0.7]{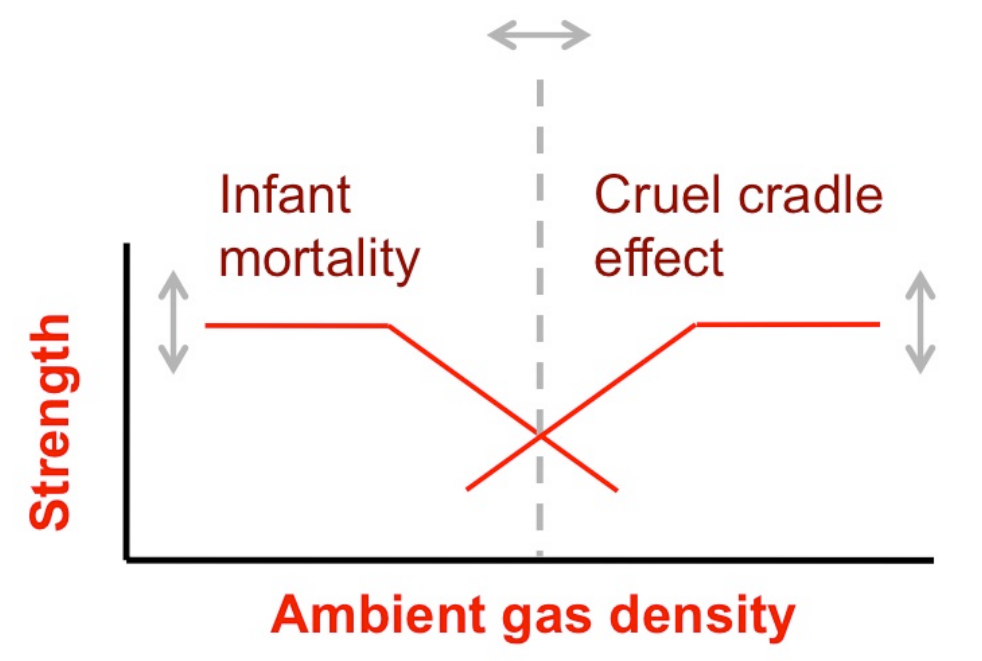}
\caption{\label{fig:cfe}
Schematic representation of the relative importance of the mechanisms that set the cluster formation efficiency (CFE). Shown are the disruptive strengths of infant mortality and the cruel cradle effect as a function of ambient gas density. The slopes, normalisations, and the location of the cross-over point of the curves are all unknown.
}
\end{figure}

Contrary to infant mortality, which should peak in low-density regions, the cruel cradle effect decreases the CFE in high-density regions. This implies that in different environments, the CFE may well be determined by different physical mechanisms. In order to understand the relation between the CFE and the gas or star formation rate density \citep[e.g.][]{goddard10,silvavilla11,adamo11}, the relative contributions of these mechanisms need to be accounted for. The `importance' of infant mortality and the cruel cradle effect is shown schematically in Fig.~\ref{fig:cfe} as a function of the ambient gas density $\rho_{\rm amb}$. While the precise form of these relations is unknown, the evidence outlined above leads to the trends sketched in the diagram. Infant mortality becomes more effective as the ambient density decreases, with the curve flattening at very low densities to reflect the saturation that occurs when star formation is so dispersed that virtually all structure would be disrupted by gas expulsion. The efficiency of the cruel cradle effect increases with ambient density, with the curve potentially flattening at very high densities where (nearly) all newly formed clusters are immediately disrupted by the strong and numerous tidal shocks. The relation between the CFE and $\rho_{\rm amb}$ is then given by the product of the curves due to infant mortality and the cruel cradle effect.

The relations in Fig.~\ref{fig:cfe} apply to the fraction of star formation that takes place in initially bound (sub)clusters, meaning that star formation in unbound associations is not included. If the fraction of star formation taking place in unbound associations also exhibits a trend with ambient gas density, it should be multiplied with the result of Fig.~\ref{fig:cfe} to obtain the actual CFE--$\rho_{\rm amb}$ relation. In the coming years, combined theoretical and observational efforts should enable the quantification of the relative contributions of star formation in unbound associations, infant mortality, and the cruel cradle effect.

\section{Using Gaia to distinguish between the mechanisms that determine the cluster formation efficiency}
\begin{figure}
\center
\includemovie[poster=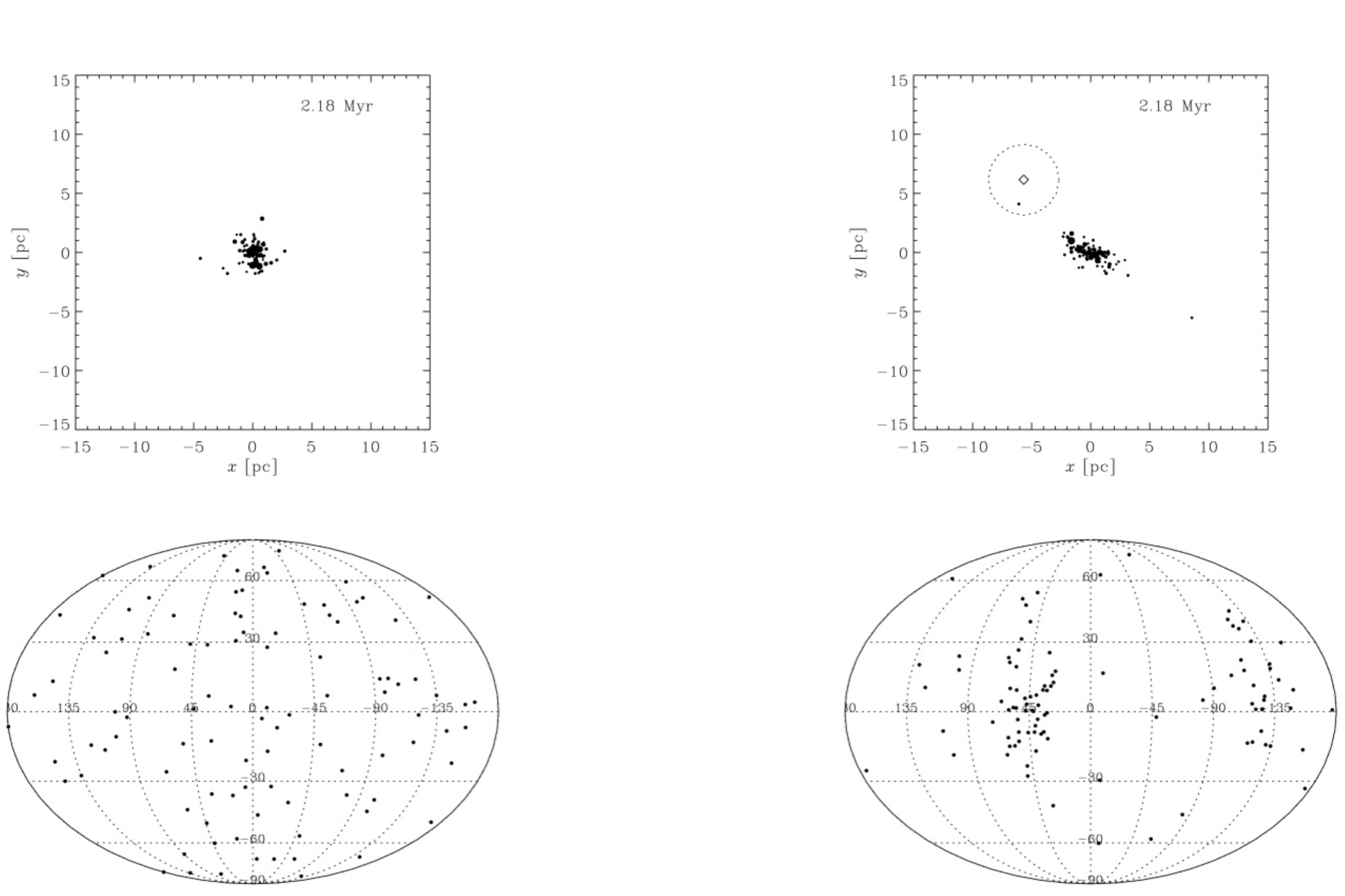,repeat]{11.475cm}{7.65cm}{Kruijssen_D_MM1.mov}
\caption{\label{fig:movie} Movie of two toy $N$-body simulations to illustrate the kinematic difference between the disruption of a $\sim 25$~\msun (sub)cluster by infant mortality ({\it left}) and the cruel cradle effect ({\it right}). Infant mortality is modelled by removing a static background potential with the same properties as the (sub)cluster at $t=3$~Myr, while the cruel cradle effect is characterised by the passage of a GMC with mass $1.25\times 10^4$~\msun. The top panels show the projected spatial configuration of the stars in their centre of mass frame. In the top-right panel, the GMC is indicated with a diamond and a dotted circle, which denotes its Plummer radius. In the bottom panels, the directions of the velocity vectors of the stars are projected onto a sphere, with angles in degrees. Infant mortality retains the random orientation of the velocities, but the cruel cradle effect induces structure in velocity space. Click on the figure to start the movie. If your PDF reader does not support embedded movies, you can also access it \href{http://www.astro.uu.nl/~kruijs/Kruijssen_D_MM1.mov}{here}.
}
\end{figure}
Because the early disruption of stellar structure due to infant mortality and the cruel cradle effect should occur on similar time scales, it is hard to distinguish between both mechanisms observationally. However, the kinematic properties of disrupted cluster remnants will depend on the process that destroyed them. When gas expulsion leads to the disruption of a young cluster, the orientations of the velocity vectors of the stars remain random -- after all, there are no external torques acting on the cluster. However, the cruel cradle effect has a different impact, since disruption by an external perturbation does lead to a preferential orientation of the stellar velocities. This is illustrated in Fig.~\ref{fig:movie}, which shows a movie of two toy $N$-body simulations in which a low-mass (sub)cluster is being disrupted. The striking difference in velocity space implies that it should be possible to distinguish between infant mortality and the cruel cradle effect. While Fig.~\ref{fig:movie} shows two highly idealised scenarios, it also indicates the key characteristics that can be used to observe in which environments infant mortality and the cruel cradle effect play a role. An additional advantage of considering the kinematics of young cluster {\it remnants} is that they need not be caught in the act of being disrupted. Depending on the local strength of the Galactic tidal field, the imprint of the disruption mechanism will remain visible for some time after the actual dispersal took place.

The combination of radial velocities from spectroscopy with proper motion measurements and membership identification by Gaia will provide the 3D positions and space velocities that are needed to reconstruct the processes shown in Fig.~\ref{fig:movie}. Until these data become available, a theoretical effort should be made to consider the kinematics of star formation in unbound associations, infant mortality and the cruel cradle effect in more detail. This will expand the current qualitative understanding of the mechanisms that set the CFE to a quantitative census, which in turn will be essential for the predictive potential of star clusters as tracers of galaxy-scale star formation.

%
%
\small  
%
\section*{Acknowledgments}   
%
The Leids Kerkhoven-Bosscha Fonds is acknowledged for supporting attendance to the meeting. I am very grateful to my collaborators for inspiring discussions and help. In particular, Thomas Maschberger, Cathie Clarke, Nick Moeckel, Nate Bastian, Ian Bonnell, and Henny Lamers have contributed greatly to the work summarised here.

\bibliographystyle{aa}
\bibliography{mnemonic,mybib}

%
%
\end{document}